\documentstyle[epsf]{elsart}
\journal{Physics Letters B}

\newcommand{\mrm}[1]{\;\mbox{\rm #1}}
\newcommand{\beq}{\begin{equation}}
\newcommand{\eeq}{\end{equation}}

\newcommand{\bea}{\begin{eqnarray}}
\newcommand{\eea}{\end{eqnarray}}

\newcommand{\rfn}[1]{(\ref{#1})}
\newcommand{\Eq}[1]{Eq.~(\ref{#1})}

\newcommand{\ea}{{\it et al.}}

\newcommand{\np}[1]{Nucl. Phys. {\bf #1}}
\newcommand{\pl}[1]{Phys. Lett. {\bf #1}}
\newcommand{\pr}[1]{Phys. Rev. {\bf #1}}
\newcommand{\prl}[1]{Phys. Rev. Lett. {\bf #1}}
\newcommand{\zp}[1]{Z. Phys. {\bf #1}}
\newcommand{\epj}[1]{Eur. Phys. J. {\bf #1}}

\newcommand{\arns}[1]{Ann. Rev. Nucl. Sci. {\bf #1}}

\def\lsim{\mathrel{\vcenter{\hbox{$<$}\nointerlineskip\hbox{$\sim$}}}}
\def\gsim{\mathrel{\vcenter{\hbox{$>$}\nointerlineskip\hbox{$\sim$}}}}

\begin{document}

\begin{frontmatter}

\hfill DESY 99-022

\hfill MZ-TH/99-03

\title{\bf Supersymmetry and CP Violating Asymmetries 
in $B_{d,s}$ Decays}
\author[a1]{Gabriela Barenboim\thanksref{gabriela}} and
\author[a2]{Martti Raidal\thanksref{martti}} 
\address[a1]{Institut f\"ur Physik, Johannes Gutenberg Universit\"at,
D-55099, Mainz, Germany}
\address[a2]{Theory Group, DESY, Notkestra{\ss}e 85, D-22603 Hamburg, Germany}
\thanks[gabriela]{E-mail: gabriela@thep.physik.uni-mainz.de}
\thanks[martti]{E-mail: raidal@mail.desy.de}

\begin{abstract}
We study possible effects of supersymmetry (SUSY) in CP asymmetries 
in non-leptonic $B_{d,s}$ decays in a variety of SUSY flavour models 
considered in literature. We use both  mass insertion and vertex mixing 
methods to calculate squark-gluino box diagrams contribution to 
$B_{d,s}$-$\bar{B}_{d,s}$ mixings. With the squark mixing parameter 
$\eta=0.22$, and with large new CP phases, it turns out that the 
CP asymmetries to be measured in upcoming B-factories, HERA-B and LHC-B, 
can be completely dominated by the SUSY contribution in almost every 
considered model. 
Discrimination between the different models can be done by comparing 
experimental results in different decay modes. In some models squark 
masses up to $\sim 5$ TeV can be probed through these experiments
provided the SUSY contribution to $B$-$\bar{B}$ mixing is at 10\% level,
$|M^{SUSY}_{12}/M_{12}^{SM}|\sim 0.1.$  This implies  
that models with heavy squarks have a fair chance to  be tested in 
the future CP experiments before LHC. 
\end{abstract}
\end{frontmatter}

\section{Introduction}

There are two major questions to be answered in particle physics.
One is the possible origin of CP violation, to be tested in 
forthcoming
B-factories and the dedicated experiments LHC-B and HERA-B. The other
one is the possible existence of supersymmetry (SUSY) as evidenced 
by the continuing efforts of both experimentalists and theorists
in searching for new physics beyond the Standard 
Model (SM). The SM has specific predictions \cite{buras}
on the size as well as 
on the patterns of CP violation in $B_{d,s}$ meson decays which,
if disproved in future experiments, would signal doubtless 
the existence of  new physics \cite{quinn}.

SUSY theories have been subject of an
 extensive study  from the  flavour physics  point of view
since they naturally offer new sources of flavour
changing neutral currents (FCNC) and CP violation, arising from the
Yukawa and SUSY breaking sectors 
of these models. These extra contributions
are already constrained in a rather stringent way 
by the experimental bounds 
on $\epsilon_K,$ $K,$ $B,$ $D$ mixings, electric dipole moments of electron
$d_e$ and neutron $d_n$ as well as on the branching ratio of 
$b\to s\gamma.$ Many works on radiatively induced FCNC processes
in SUSY \cite{early} have 
concentrated on the minimal supersymmetric
extension of the SM (MSSM), and  assumed
 often the minimal supergravity (mSUGRA)  
framework for the SUSY breaking sector. Updated analyses 
of mSUGRA imply \cite{msugra}
that  $\Delta M_{B_d},$ $\epsilon_K$ as well as  the branching ratio of the 
processes $b\to s ll$ may be enhanced at most by a  few tens of percent
while the branching ratios of $b\to s\nu\bar\nu$ and  
$K\to \pi\nu\bar\nu$ may be reduced by only a 
few percent. The new SUSY
CP phases have negligible effects on the decays
$B\to X_s\gamma$ and  $B\to X_s ll$ in mSUGRA. In particular, there is no
new large phase shift in the $B$-$\bar B$ mixing \cite{msugracp} implying
no major deviations in any CP asymmetry as compared with the
SM predictions.

\setcounter{page}{1}

This conclusion changes dramatically as soon as one considers
SUSY GUTs or other SUSY models without universality in the soft breaking
sector. As the squark mass matrices are in general independent of 
the quark mass matrices, they introduce a major source of FCNC which, 
together with new CP phases, can substantially modify the SM 
predictions for 
CP violating observables to be measured in B-factories and LHC-B.   
There is a plethora of SUSY models, to be considered below, which satisfy
all the present phenomenological constraints and allow large 
signals of new physics in these experiments\footnote{For a review of
SUSY CP violating signals at future colliders see Ref. \cite{bkrw}.}.

Many of the recent works on SUSY signatures in CP violating
decays of $B_{d,s}$ mesons have concentrated 
on studying   new physics contributions to decay amplitudes. 
These analyses cover the
full spectrum of SUSY models. Large deviations from the SM in
CP asymmetries in $b\to s\gamma$ are predicted in Ref. \cite{bsg}
and in penguin dominated non-leptonic $b$ decays  in Ref. \cite{decay}.
As the new physics contributions to decay amplitudes are non-universal,
comparison of CP asymmetries in different decay modes allows 
to find new physics in a model independent way.
Surprisingly, while the effects of new physics contribution 
to $B$-$\bar B$ mixing in CP violating {\it lepton} 
asymmetries have been analyzed
in almost all possible SUSY flavour models in both $B_d,$ $B_s$ 
decays \cite{randall},  the well known mixing effects in
CP asymmetries in {\it non-leptonic} $B_d,$ $B_s$ decays have been considered
in detail only in a few particular models 
of supersymmetry \cite{cohen,barbieri1}. 
The mass insertion method has been used to describe the internal 
squark effects in these works.     
These  recent analyses  have shown  \cite{grs} that LEP2 experiments 
pose a serious naturalness problem to all conventional  mSUGRA models, 
motivating the search for different SUSY models.
In addition,
motivated by the recent claim \cite{oscar} that the CP violating
observable $\epsilon_K$ may have a fully supersymmetric
origin in general SUSY models, we feel encouraged to study the predictions of
the full range of SUSY flavour models considered in literature 
on the CP asymmetries to be measured in the upcoming experiments.   

Following the model independent analyses of Ref. \cite{sanda}
we study in this letter  the CP violating asymmetries in $B_{d,s}$ 
meson decays in the SUSY flavour models classified in Ref. \cite{randall}.
This includes models where FCNC and CP problems are solved using
alignment \cite{lns1,ns,nr,chmoroi},
non-abelian symmetries 
\cite{chmoroi,bdh,bhrr,chmu1,chmu2,pt,dlk,ps,ks}
and decoupling of squarks of the first two generations 
\cite{pt,dg,dp,ckn,mr,nw,an}.
These cover a large variety of general SUSY flavour models considered
in literature.  We use both the mass insertion method \cite{mi,ggms}
as well as the vertex mixing method \cite{vm} to describe the dominant
gluino-squark box diagram's contribution  to the  
meson mixing. We consider both $B_d$ and $B_s$ decays, as they
provide complementary information for distinguishing new physics from
the SM as well as discriminating between different SUSY models. 
We show that despite  the quite strong bounds on the squark
masses in these models, given the predictions for squark mixings, 
the SUSY CP violating contribution in $B_{d,s}$ systems might 
dominate over the SM one in almost all the models.  
In some models measurable effects can be achieved for
squarks masses as high as several TeV.
This implies the
possibility of discovering SUSY in B-factories before LHC.

\section{New physics in $B$-$\bar B$ mixing}

Possible ways of evidencing 
the existence of new physics in B-factory experiments have been 
studied in the literature in a model independent way.   
Signals of new physics arising form new contributions in 
the decay amplitudes 
have been studied in \cite{decay} and are  beyond the scope of this paper.
If new physics contributes to the $B$-$\bar B$ mixing, as it is the 
case with the general SUSY models, it has been shown,  
that the planned experiments will be able to distinguish the effects 
of new CP phases from the effects of the SM 
 Cabibbo-Kobayashi-Maskawa (CKM) phase \cite{akl}.
This can be done by comparing the CP asymmetries in the different
decay modes of both $B_d$ and $B_s$ mesons. This important 
result implies that future experiments will allow us to  discriminate
between different SUSY models, as will be shown below.

In terms of the off-diagonal elements of the
$2\times 2$ $B^0_q$-$\bar{B}^0_q,$ $q=d,\,s,$ 
mixing Hamiltonian ${\bf M} -{\rm i}
{\bf \Gamma}/2$   the $B^{0}_{q}$-$\bar B^{0}_{q}$ mixing phase 
$\phi_M^{q}$ is expressed as
\bea \label{lam}
e^{-2i\phi_M^{q}}= \sqrt{\frac{M_{12}^{q*}-\frac{i}{2}\Gamma_{12}^{q*}}
{M_{12}^{q}-\frac{i}{2}\Gamma_{12}^{q}}}\,.
\label{lambda}
\eea
In the analysis that follows, we make use of the fact that 
$M_{12}$ can be parametrized as \cite{sanda}
\bea
M_{12}=M_{12}^{SM}+M_{12}^{SUSY}=M_{12}^{SM}
\left(1+h e^{i\theta}\right)\,,
\eea
where $M_{12}^{SUSY}$ denotes the new SUSY contribution 
and $h=|M_{12}^{SUSY}/M_{12}^{SM}|.$ The relative phase  
$\theta=\mrm{arg}(M_{12}^{SUSY}/M_{12}^{SM})$ is in principle 
not restricted because it results from  
the possibly large new phases arising in the squark sector.
As $|\Gamma_{12}|\ll |M_{12}|$ and 
 $\Gamma_{12}\approx\Gamma_{12}^{SM},$  the  $B^{0}$-$\bar B^{0}$ mixing 
gets modified by the phase $\phi_M$ 
\bea
\phi_M=\arctan\left( \frac{h\sin\theta}{1+h\cos\theta}\right)\,.
\label{phi}
\eea
In the $B_d$ system the new physics contribution is constrained by the 
measured $B_H^{0}$-$B_L^{0}$ mass difference which may by expressed as 
$\Delta M_{B_d}=2|M_{12}^d|.$

The phase \rfn{phi} directly modifies the CP asymmetries as discussed 
in \cite{akl,sanda}. Its effects are universal in all decay modes.
Because the SM predicts almost vanishing CP asymmetries in most of
$B_s$ decays, non-zero experimental results in these measurements
unambiguously measure the new physics phase $\phi_M.$ However, in 
$B_d$ system (to be experimentally probed in B-factories) the CP asymmetries
can be also large within the SM, therefore finding new physics 
contributions
to $B$-$\bar B$ mixing requires an accurate knowledge of the SM
predictions for these asymmetries. 
In the presence of new phases, the CP asymmetries which measure the 
SM CKM angle $\beta$ (e.g. in $B_d\to J/\psi K$) are modified as 
$a_{CP}=-\sin 2(\beta+\phi_M)$ while  the CP asymmetries which measure the 
SM CKM angle $\alpha$ (e.g. in $B_d\to \pi^+\pi^-$) receive a
new contribution with the opposite sign, $a_{CP}=-\sin 2(\alpha-\phi_M).$
As the new phase, $\phi_M$, cancels out in $\alpha + \beta$ measurements
of the third CKM angle $\gamma$ are crucial.

However, if the new physics contribution is large enough so that the 
measured CP asymmetries are outside the allowed SM regions then, the 
new physics can be traced off unambiguously also in B-factories.
Fortunately, recent global analyses \cite{al}
show that the precision reached in constraining the CKM matrix is 
quite good even  now, yielding the result $\sin 2\beta=0.73\pm 0.21$
at 95\% confidence level.  Therefore it follows from \Eq{phi} 
that a SUSY contribution of $h=0.1$, together with large phase $\theta$,
implies measurable deviations from the SM. In the following we assume
that the minimal detectable value of $h$ is 0.1. Note that this assumption
is somehow 
conservative, with better experimental precision, in the future
much smaller effects can be probed.

\section{SUSY flavour models}

There are two distinct sources of flavour violation in general SUSY models
in addition to the usual CKM mixing in the quark sector: (i) flavour
violating interactions of top quark with charged Higgs, and (ii) 
misalignment
between  the  fermion and sfermion mass matrices. The former
possibility has been studied in \cite{early} and it
 is quite constrained to lead 
to significant deviations from the SM. In this work we concentrate on the 
latter possibility. In this case the new physics contribution to the
$B_{d,s}$ mixing is dominated by the box diagrams with gluinos, $\tilde g,$ 
and squarks, $\tilde q,$ running in the loop. The new flavour mixings 
in the $6\times 6$ down squark mass matrix,
\bea
\tilde M^{d2}=\left( 
\begin{array}{cc}
\tilde{M}_{LL}^{d2}&\tilde{M}_{LR}^{d2}\\ 
\tilde{M}_{RL}^{d2}&\tilde{M}_{RR}^{d2} 
\end{array} 
\right)\,,
\label{msq}
\eea
lead to new flavour changing interactions, which,
together with the new CP phases, may significantly contribute to 
$B_{d,s}$ meson mixings. In \Eq{msq}, the phenomenological constraints on
the off-diagonal squark mixings in  $3\times 3$ matrices 
$\tilde M^{d2}_{LR,RL}$ are more stringent than in  
$\tilde M^{d2}_{LL,RR}$ \cite{ggms},  and we 
therefore neglect $\tilde M^{d2}_{LR,RL}$ 
in the following.

The most oftenly used parameterization of squark mixing effects is called the 
mass insertion method \cite{mi,ggms}.
It assumes a flavour diagonal $\tilde g q\tilde q$ vertex and 
also flavour diagonal quark mass
matrices and places all the mixing effects, described by the
dimensionless parameter $(\delta_{ij})_{MN},$ $M,N=L,\,R,$ 
(see Eq. (2.35) in Ref. \cite{randall}), in  the 
squark propagators.
This method is the appropriate one when the squark masses are 
nearly degenerate but the price to be paid for using it, is the
introduction of an 
average squark mass which should be chosen in an appropriate way \cite{ggms}.
The second method, the vertex mixing method, deals  with 
the mass eigenstates of quarks and squarks with off-diagonal
$\tilde g q\tilde q$ couplings, and considers only the contribution
from the lightest squark generation. 
Thus the mixing effects in this formalism can be characterized by  
the matrices  $K^d_L=V^d_L \tilde V^d_L$ and $K^d_R=V^d_R\tilde V^d_R,$
where $V^d_{L,R}$ and $\tilde V^d_{L,R}$ are the matrices diagonalizing
the quark and squark mass matrices, respectively.
Obviously, this is a better 
approximation when one generation is much lighter than the others.
In our numerical calculations we will use both methods depending 
upon which one is the most appropriate for the concrete application 
to the case at stake.

The SUSY flavour models were already 
classified and studied in a comprehensive work, \cite{randall}.
There are three basic mechanisms to suppress the
dangerous FCNC. The first one is the alignment of squark mass matrices 
along to the quark ones, 
so that the mixing matrices  $K^d_{L,R}$
are close to unity \cite{lns1,ns,nr,chmoroi}. 
Motivated by the different behavior of the third family as compared to
the first two ones, there have been proposed 
 models with non-abelian flavour symmetries
\cite{chmoroi,bdh,bhrr,chmu1,chmu2,pt,dlk,ps,ks}.
The approximate flavour symmetries are broken by a small factor $\eta$
which in our numerical estimates is taken to be equal to the Wolfenstein 
parameter, $\eta=0.22.$
Finally, there are models with super-heavy squarks in the first two 
generations  \cite{pt,dg,dp,ckn,mr,nw,an}. 
To suppress FCNC completely, several of the methods described above
may be combined in a single realistic model.  
It has been pointed out that in order to avoid fine tuning in the
electroweak symmetry breaking \cite{dg}, and to ensure the positivity of 
the stop mass matrix squared at weak scale \cite{ahm},
there should be bounds on the squark mass parameters in consistent
heavy squarks  models.
These, in turn, constrain the most popular scenarios of generating the
multi TeV masses for the first two generation squarks using horizontal 
$U(1)$ symmetries \cite{dp,mr,nw}. 
On the other hand, recently it has been proposed that all the SUSY soft masses
might be at multi TeV scale and  the lightness of the third 
generation squarks should be then generated radiatively \cite{fkp}.
In this scenario the right bottom squark remains heavy with mass of several
TeV. The analysis of such a model
is  beyond the scope of 
the present work.
We shall take a purely phenomenological approach and allow squark masses,
in particular sbottom masses, to vary from ${\cal O}(100)$ GeV 
to several TeV and,
study the sensitivity of the CP asymmetries to the squark masses.

It is important to notice that from
all the models listed above, some either are not able to 
pass all the phenomenological constraints
set by $\epsilon_K,$ $\Delta M_K,$  $d_e$ and $d_n$ or 
do not  have specific prediction for the squark mass matrix textures
(for discussion see Ref. \cite{randall}).
The models which have interesting predictions for 
B-physics, are summarized in Table 1. 
In the first column of this table,  we specify the appropriate method  
for the estimation of  the SUSY box contribution in each model, while
in the second column, we specify the mechanism used to suppress FCNC. 
For a detailed description of each of these models, we refer the
reader to the original literature which 
is listed in the third column. 
The predictions for the (23) and (13) flavour mixings in the 
mass matrices  $\tilde M^d_{LL,RR}$  are also given for each model.
\begin{table}
\begin{tabular}{|c|c|l|cc|c|c|cc|c|} \hline
&&&\multicolumn{4}{c}{(13) mixing}\vline&\multicolumn{3}{c}{(23)
mixing }\vline\\\cline{4-10}
&&Model&$LL$&$RR$&$m_{\tilde q}$ (GeV)&$h_{max}$&$LL$&$RR$&$ h_{max}$ \\ \hline
&&\cite{lns1}&
$\eta^3$&$\eta^3$& 550 & 1.2 &
$\eta^{2}$&$\eta^{4}$&  0.016 \\
&A&\cite{ns}, \cite{chmoroi} a&
\multicolumn{2}{c}{Too small}\vline & 1400 & --- &
$\eta^2$&1&  1.05\\ 
Mass&&\cite{nr} &
$\eta^{3}$& $\eta^{7}$& 260 & 0.18 &
\multicolumn{3}{c}{Small CP angle}\vline \\\cline{2-10}
inser-&&\cite{chmoroi} b&$\eta^3$&$\eta^{3/2}$& 1400& 1.3 &
$\eta^{2}$&$\eta^{1/2}$& 1.05\\ 
tion&B&\cite{bhrr}, \cite{pt} b
&$\eta^3$&$\eta^3$& 550& 1.2 &
$\eta^{2}$&$\eta^{2}$& 0.96\\
&&\cite{chmu1}&$\eta^2$&$\eta^4$&330 &1.1 &
$\eta^{3}$&$\eta^{5}$& 0.002\\
&&\cite{ps}&$\eta^3$&$\eta^4$&260 &1.2 &
$\eta^2$&$\eta^4$& 0.07 \\  \hline
Vertex&B$+$C&\cite{pt} a
&$\eta^3$&$\eta^3$& 830& 1.3&
$\eta^{2}$&$\eta^{2}$& 0.97\\ \cline{2-10}
mixing&C&\cite{ckn},\cite{nw} &$\eta^3$& & 450& 1.2 &
$\eta^2$& & 0.89 \\\hline
\end{tabular}
\caption{
SUSY models which  solve the FCNC problem with A--alignment, 
B--non-abelian symmetries and C--heavy squarks in first two generations.
The mixing parameter $\eta$ in $LL$ and $RR$ squark mass matrices is taken 
to be $\eta=0.22$ in our numerical estimates. 
The lower bounds on $m_{\tilde q}$ coming from the measurement of 
$\Delta M_{B_d}$ are presented for each model. The maximal values of 
$h=|M_{12}^{SUSY}/M_{12}^{SM}|$ in $B_d$ and $B_s$ systems for each model
are calculated for the given lower bound on $m_{\tilde q}.$
}
\end{table}

\section{Numerical results}

In the SM the quantity $M_{12}^{SM}$ has been calculated including NLO QCD
corrections (for references see \cite{buras}) and reads
\begin{equation}
M_{12}^{ SM}=\frac{G_{F}^2}{12\pi^2}\eta_{ QCD}B_{B_{q}}f_{B_{q}}^2 
M_{B_{q}}M_W^2(V_{tq}V_{tb}^*)^2 S_0(z_t),
\label{smm}
\end{equation}
where 
\begin{equation}
 S_0(z_t)=\frac{4z_t-11z_t^2+z_t^3}{4(1-z_t)^2}
-\frac{3z_t^3\ln z_t}{2(1-z_t)^3}\,,
\end{equation}
with  $z_t=m_t^2/m_W^2.$
Recent updated  values for the bag parameter,  $B_{B_d}$, and the decay 
constant, $f_{B_d}$, are  $B_{B_d}=1.29\pm{0.08}\pm{0.06}$ and
$f_{B_d}=175\pm{25}$ MeV, respectively, and  the QCD
correction factor $\eta_{ QCD}$ takes the  value 
$\eta_{ QCD}=0.55\pm0.01$ \cite{buras}. 
The meson masses we have used for the numerical
estimation are,
$M_{B_d}=5.279$ GeV,  $M_{B_s}=5.369$ GeV, and the value of the
top quark mass, $m_t(m_t)=165$ GeV, is taken from \cite{buras}.
For the CKM mixing elements we use the following values
$|V_{td}V^*_{tb}|=0.0084$ and $|V_{ts}V^*_{tb}|=0.040.$

The dominant contribution to $M_{12}^{ SUSY}$ comes from the $\Delta B=2$
box diagrams with $\tilde q,$ $\tilde g$ running in the loop.
This  can be calculated either in the
scenario of vertex mixing \cite{vm} or using mass insertion
\cite{mi,ggms}.  The latter method is widely used in literature
while the former, which should be more suitable for models with super heavy 
squarks, is not. 
The LO \cite{lo} and NLO \cite{nlo} QCD corrections to $\Delta F=2$ 
processes in SUSY models are calculated using mass insertion method.
There are no calculations of QCD corrections using the vertex mixing
method. Since one of the aims of this paper is to compare 
the results in vertex mixing and mass insertion we neglect QCD corrections 
here and work at the level of electroweak box diagrams.
Because the QCD corrections are known to {\it enhance} the SUSY contribution
significantly (for the Wilson coefficients using mass insertion 
see, e.g., Ref. \cite{barbieri1}), 
the results in this work should be regarded as a 
{\it conservative} estimate of the
SUSY contribution to CP asymmetries. 
Moreover, as what matters in CP asymmetries are the ratios of amplitudes, 
QCD corrections tend to  cancel.

Within  the vertex mixing approximation,
 $M_{12}^{ SUSY}$  is given by \cite{vm}  
\begin{eqnarray}
\lefteqn{M_{12}^{
VM}=-\frac{\alpha_s^2}{216m_{\tilde{q}}^2}
\frac{1}{3}B_{B_{q}}f_{B_{q}}^2M_{B_{q}}
\left\{((K_{L}^d)_{3i}^2+(K_{R}^d)_{3i}^2)
(66\tilde{f}_4(x)+24xf_4(x))+\right.}
\label{cohen}\\
&&\hspace*{-0.4cm}\left.(K_{L}^d)_{3i}(K_{R}^d)_{3i}
\left[\left(36-24\left(\frac{M_{B_{q}}}{m_b+m_q}\right)^2\right)
\tilde{f}_4(x)+\left(72+384\left(\frac{M_{B_{q}}}{m_b+m_q}\right)^2
\right)xf_{4}(x)\right]\right\},\nonumber\\
&&f_4(x)=\frac{2-2x+(1+x){\rm ln}x}{(x-1)^3},\hspace{0.5cm}
\tilde{f}_4(x)=\frac{1-x^2+2x{\rm ln}x}{(x-1)^3},
\end{eqnarray}
where $x=m_{\tilde{g}}^2/m_{\tilde{q}}^2,$
$i=1,2$ for $B_d$ and $B_s,$ respectively and we take 
 $(K_{L,R}^d)_{33}\sim{1}$.
Within the mass insertion notation, $M_{12}^{ SUSY}$  takes the form
\cite{ggms}  
\begin{eqnarray}
\lefteqn{M_{12}^{
MI}=-\frac{\alpha_s^2}{216m_{\tilde{q}}^2}
\frac{1}{3}B_{B_{q}}f_{B_{q}}^2M_{B_{q}}
\left\{((\delta_{3i})_{LL}^2+(\delta_{3i})_{RR}^2)
(66\tilde{f}_6(x)+24xf_6(x))+\right.}\label{ma}
\\
&&\hspace*{-0.2cm}\left.(\delta_{3i})_{LL}(\delta_{3i})_{RR}
\left[\left(36-24\left(\frac{M_{B_{q}}}{m_b+m_q}\right)^2\right)
\tilde{f}_6(x)+\left(72+384\left(\frac{M_{B_{q}}}{m_b+m_q}\right)^2
\right)xf_{6}(x)\right]\right\},\nonumber\\
&&\hspace*{-0.2cm}f_6(x)=\frac{6(1+3x){\rm ln}x+x^3-9x^2-9x+17}{6(x-1)^5},
\\
&&\hspace*{-0.2cm}
\tilde{f}_6(x)=\frac{6x(1+x){\rm ln}x-x^3-9x^2+9x+1}{3(x-1)^5}.
\end{eqnarray}
For our numerical estimates we use $\alpha_s(m_b)=0.222,$ 
$m_b(m_b)=4.4$ GeV and  $x=m_{\tilde{g}}^2/m_{\tilde{q}}^2=1,$
unless stated otherwise.

To begin with, 
we have to take into account the constraint on the squark masses 
coming from the measurement $\Delta M_{B_d}=0.470\pm 0.019$ ps$^{-1}.$ 
In order to do so we require 
that our calculated $\Delta M_{B_d}=2|M_{12}^d|$  coincides, 
within errors, with the experimental value. The errors are completely 
dominated by the errors of $B_{B_d}$ and $f_{B_d}.$
The lower bounds on the squark masses, $m_{\tilde q}$, 
for $\eta=0.22$ and for a fixed value of $\theta$,
$\theta=\pi/2$, (that  takes into account  the case 
where the  SUSY CP violation is large)
for each model are presented in Table 1. The interpretation of each of
these numbers depends on the method used to calculate the SUSY contribution.
For the mass insertion method (squark masses are nearly degenerate) 
this is the average squark mass, while for the vertex mixing method
(squarks of the first two generation are decoupled) it bounds 
the sbottom masses.

In principle, one can get bounds on the squark masses and mixings also
from the measurement of $b\to s\gamma$ branching ratio. However, as we
neglect the $LR$ mixings in \Eq{msq} here, the bound on the (23) squark 
mixing in $LL$ is at best of order ${\cal O}(1)$ for $m_{\tilde q}=500$
GeV \cite{ggms}, even assuming that the 
SUSY contribution to the branching
ratio is less than 10\%. Comparing this bound with the model predictions in
Table 1 we conclude that $b\to s\gamma$ does not impose constraints 
on our results.

Comparing the squark mass bounds in Table 1, it follows that 
in models with the same $LL$ and $RR$ mixings the 
vertex mixing method gives much larger contribution to $M_{12}$ 
than the mass insertion one 
(compare models \cite{lns1}, \cite{bhrr}, \cite{pt} b with \cite{pt} a).   
Since the models we have considered here
have very different predictions for the squark
mixings the mass bounds vary from the direct Tevatron bound
 $m_{\tilde q}\gsim 260$ GeV to 1.4 TeV. 

For the given squark mass bounds, the maximum allowed values of 
$h=|M^{SUSY}_{12}/M^{SM}_{12}|$ for both $B_{d,s}$ systems in each 
model are also given in Table 1. For the models \cite{ns} and 
\cite{chmoroi} a, as there is no bound coming from 
$\Delta M_{B_d}$  and in order to be conservative, 
we give $h$ in the $B_s$ system for 
$m_{\tilde q}=1.4$ TeV.
As it is evident from the results shown in the table,
the SUSY contribution dominates in all the models 
besides model \cite{nr}, where it can be still measurable in the $B_d$ system. 
As each model predicts   different squark mixings
 the maximum $h$ is different in  the $B_{d}$ and $B_{s}$ systems.

\begin{figure}[t]
\centerline{
\epsfxsize = 0.5\textwidth \epsffile{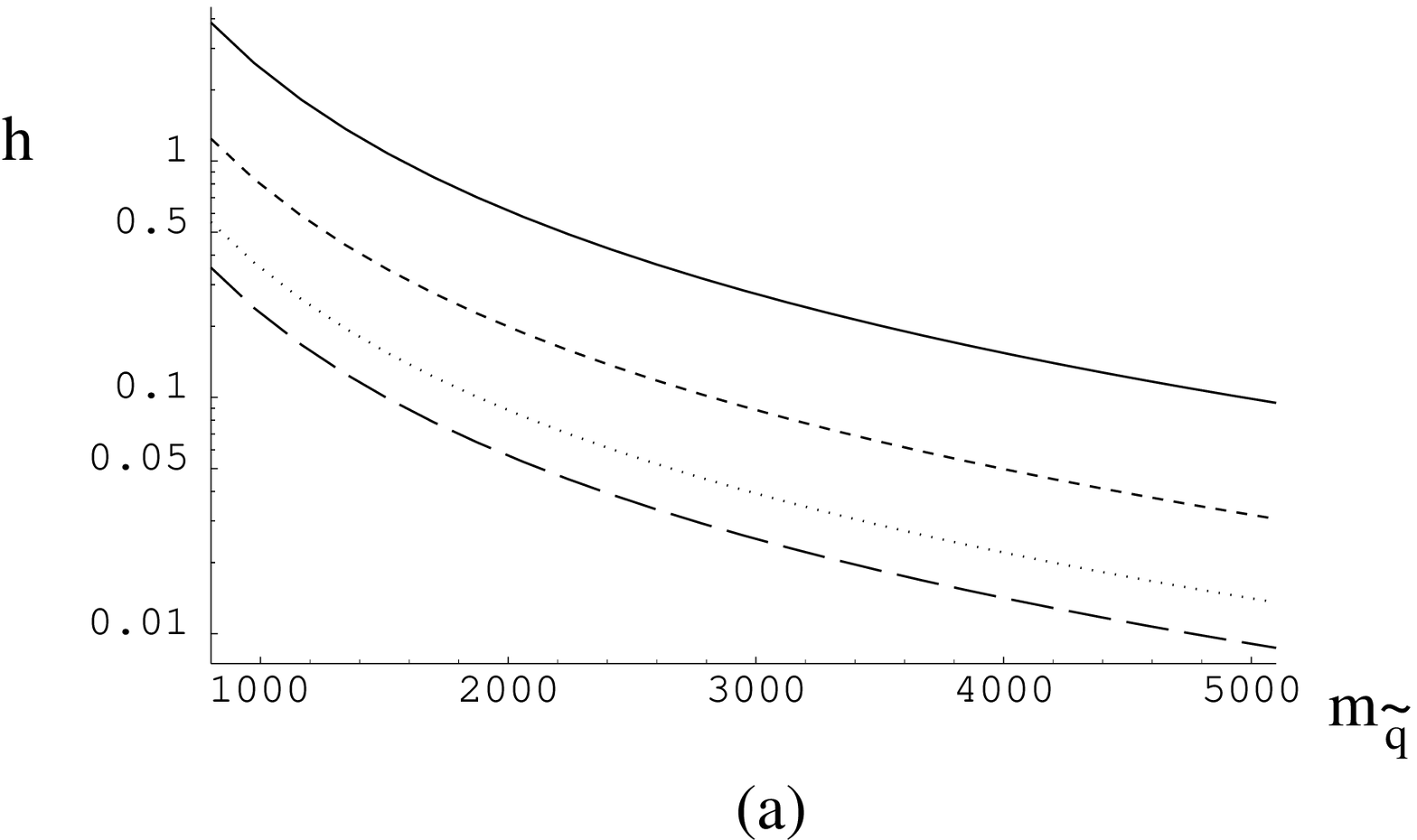} 
\hfill
\epsfxsize = 0.5\textwidth \epsffile{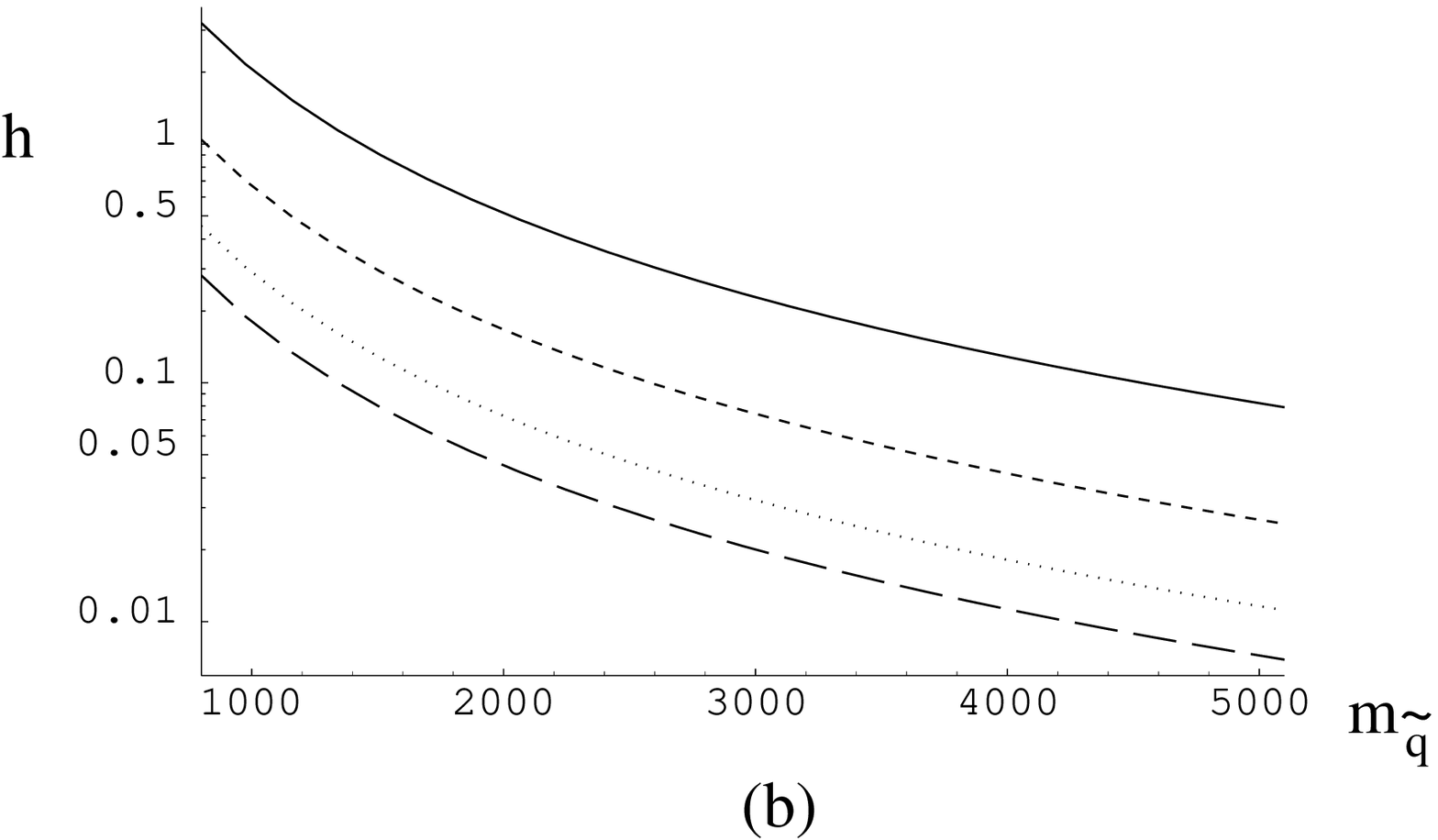}
}
\caption{ $h=|M_{12}^{SUSY}/M_{12}^{SM}|$ 
as a function of  $m_{\tilde{q}}$ (in GeV) for $x=1$ for $B_d$ (figure (a))
 and $B_s$ (figure (b)) systems. 
The solid curves are for the model \cite{chmoroi} b (figure (a))  and 
for  the models \cite{ns}, \cite{chmoroi} a,b  (figure (b)).
The dotted curves denote always models \cite{bhrr},  \cite{pt} b,
the long dashed curves models \cite{ckn}, \cite{nw}, and 
the short dashed curves model \cite{pt} a.
}
\label{h}
\end{figure}

The largest SUSY contribution to the $B_d$-$\bar B_d$ mixing is 
predicted within models \cite{chmoroi} b and \cite{pt} a, 
and that to the $B_s$-$\bar B_s$ mixing by the models 
\cite{ns}, \cite{chmoroi} a and b, and \cite{pt} a. 
To study  how large squark masses can possibly be probed in the prospective
CP experiments we plot in Fig. \ref{h} the dependence of $h$ on $m_{\tilde q}$
in these models ($B_d$ in Fig. 1 (a) and $B_s$ Fig. 1 (b)). 
The solid curve in Fig. 1 (a) is for  model \cite{chmoroi} b  
while in Fig. 1 (b) it is for all the models in \cite{ns,chmoroi} 
because their predictions coincide. 
The dotted curves denote always models \cite{bhrr},  \cite{pt} b,
the long dashed curves models \cite{ckn}, \cite{nw}, and 
the short dashed curves model \cite{pt} a.
Numerically $h$ is sizable,
i.e. it has values of  order 0.1, up to 
squark masses of several TeV as can be seen from the graphs.

\begin{figure}[t]
\centerline{
\epsfxsize = 0.5\textwidth \epsffile{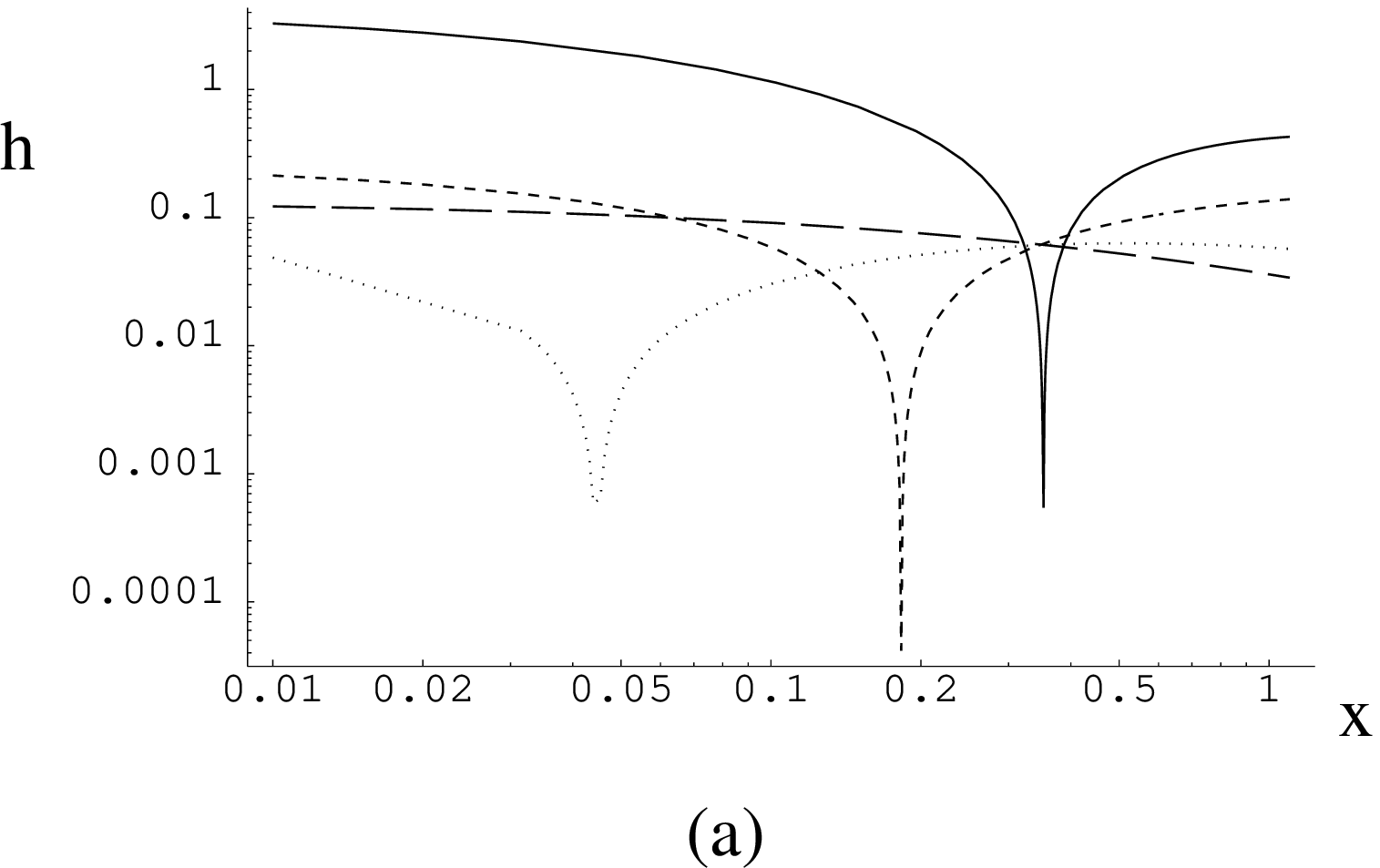} 
\hfill
\epsfxsize = 0.5\textwidth \epsffile{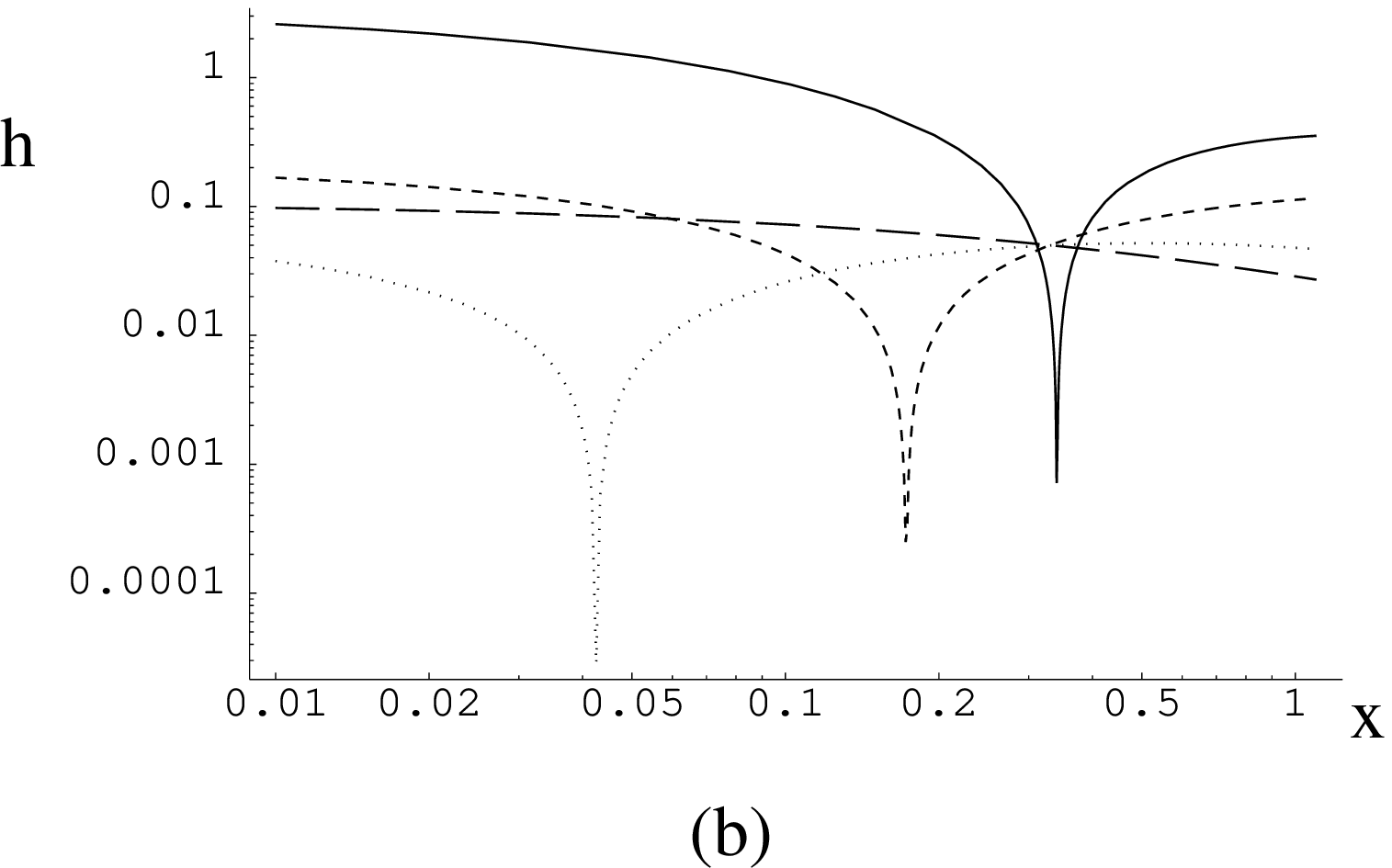}
}
\caption{ $h=|M_{12}^{SUSY}/M_{12}^{SM}|$ 
as a function of  $x=m_{\tilde{g}}/m_{\tilde{q}}$  
for $m_{\tilde q}=2.5$ TeV for $B_d$ (figure (a))
 and $B_s$ (figure (b)) systems. 
The notation is the same as in Fig. \ref{h}.
}
\label{x}
\end{figure}

To account for the case with very heavy squarks we plot, 
in Fig. \ref{x},
$h$ as a function of $x$ for a fixed value $m_{\tilde q}=2.5$ TeV.
The models and notation are  the same as in Fig. \ref{h}.
Because of the 
large squark mixings, $h$ is always largest in 
model \cite{chmoroi} b. However, in models where the
first two squark
generations are decoupled (models \cite{pt} a, \cite{ckn,nw} )
and where the vertex mixing method is used, the value of $h$ is in
general increased for small values of $x.$
Therefore, models with large sbottom masses (gluinos are expected to be light)
can still be tested in future experiments.

\section{Discussion and conclusions}

As discussed in Section 2,
the  measurements of non-leptonic  CP asymmetries 
through different $B_d$ and $B_s$ decay modes, allow one 
to discriminate
between new physics and the SM  description of CP violation arising solely
from the CKM phase. Our 
studies show that these measurements can also discriminate between the 
different SUSY models.  
According to \Eq{phi}, it follows from 
Table 1, that SUSY can completely dominate 
the CP asymmetries to be measured in B-factories and LHC-B. This is 
so even once the very strong bounds on the squark masses in Table 1
are properly taken into account.
Large effects in non-leptonic CP asymmetries in $B_d$ system can be 
expected almost 
for any prediction of the relative size of the $LL$ and $RR$ mixings. 
This should be compared with the leptonic asymmetries where new physics 
can be probed only if $LL=RR$ \cite{randall}.
As the different models, in general, 
have different predictions for the 
(13) and (23) mixings of the squark mass matrices (which may differ from the
CKM mixing structure),  the comparison of the 
CP asymmetries in $B_d$ and $B_s$  allows a clear discrimination 
between models.
For example, models \cite{ns} and \cite{chmoroi} a can give large 
contributions
 to CP asymmetries only in $B_s$ decays while the rest of 
the models can modify either $B_d$ or both mesons decays.

It is also instructive to compare the mass insertion method with the 
vertex mixing one. It follows from Table 1 that in models with super heavy 
squarks in 
the first two generations, the SUSY contribution for the same
value of $m_{\tilde q}$ is larger than in other models. This is because
 in models with nearly degenerate   squarks, the GIM mechanism is operative
while in heavy squark models, only the 
$b$ squarks contribute.  
Also, the results in the vertex mixing case are much more 
independent of the relative magnitude of $LL$ and $RR$ mixings
(compare, for example, models \cite{ckn,nw} with models \cite{chmu1,ps})
implying that the new physics should affect the CP asymmetries 
in super heavy squark models more strongly.

Finally, it  remains to be answered how large the squark masses can
be and  still give  measurable effects in CP asymmetries. 
This question becomes very interesting in light of the results of
Ref. \cite{grs}. These authors claim that LEP2 results may indicate that 
colored particles have masses of few TeV.
Even in a very disadvantageous 
scenario, where the SUSY effects can be 
as low as 10\%, i.e. $h=0.1$,  but provided that the
phase $\theta$ is large enough, there will be still observable 
effects in upcoming B-factories as shown in Section 2. 
As can be seen from Fig. 1 (a), $m_{\tilde q}\lsim 5.1$ TeV and 
$m_{\tilde b}\lsim 3$ TeV in models \cite{chmoroi} b and \cite{pt} a,
respectively, can be probed for $\eta=0.22.$ 
The $B_s$ decays (Fig. 1 (b)) are 
somehow less  sensitive, implying 
 $m_{\tilde q}\lsim 4.5$ TeV and 
$m_{\tilde b}\lsim 2.6$ TeV in models \cite{ns},
\cite{chmoroi} a,b and \cite{pt} a, respectively.
If the sensitivity of the future experiments would turn out to 
be better than 10\%
then, higher masses can possibly be probed.
For smaller values of $\eta$ the mass reach scales linearly. 
These results have important and far reaching 
implications on the heavy squark models.
As discussed, in some models the right bottom squarks might have
masses of several TeV. Have the new large SUSY CP phases happen 
in this sector, 
our results would imply (plots in Fig. \ref{x}) that TeV masses 
can still give observable effects in B-factories.

In conclusion, we have shown that, with the squark mixing parameter 
$\eta=0.22$
and with large new SUSY CP phases, the CP asymmetry measurements in upcoming 
B-factories, HERA-B and LHC-B can be completely dominated by the 
SUSY contribution in almost every SUSY flavour model that 
we have considered. 
Discrimination between the different models can be done by comparing 
the results of the  different decay modes.
Assuming that SUSY effects  at the level of 10\% are still measurable, 
namely a $|M^{SUSY}_{12}/M_{12}^{SM}|\sim 0.1$ can be tested,
in some models the sensitivity is enough to explore squark masses up 
to $\sim 5$ TeV. This implies that the 
models with heavy squarks have a great chance of being tested 
in future experiments.

\begin{ack}
We thank A. Ali for discussions and reading the manuscript.
MR is grateful to A. von Humboldt Foundation for the grant
whereas G.B. acknowledges
a post-doctoral fellowship
of the Graduiertenkolleg ``Elementarteilchenphysik bei 
mittleren und hohen Energien"
of the University of Mainz.
\end{ack}


\begin{thebibliography}{99}


\bibitem{buras}
A.J. Buras and R. Fleischer,  hep-ph/9704376,
in {\em Heavy Flavours II}, World Scientific (1997), 
eds. A.J. Buras and M. Linder. 

\bibitem{quinn} Y. Nir and H.R. Quinn, \arns{42},  211 (1992);
M. Gronau and D. London, \pr{D 55}, 2845 (1997);
Y. Grossman, Y. Nir and R. Rattazi, SLAC-PUB-7379, hep-ph/9701231.


\bibitem{early} See, e.g., 
S. Bertolini, F. Borzumati, A. Masiero and G. Ridolfi,
\np{B353} (1991) 591;
N. Oshimo, \np{B404} (1993) 20;
J. Hewett, \prl{70} (1993) 1045;
R. Barbieri and G. Giudice, \pl{B309} (1993) 86;
F.M. Borzumati, \zp{C63} (1994) 291;
A. Ali, G. Giudice and T. Mannel, \zp{C67} (1995) 417;
S. Bertolini and F. Vissani, \zp{C67} (1995) 513;
P. Cho, M. Misiak and D. Wyler, \pr{D54} (1996) 3329;
H. Baer and M. Brhlik, \pr{D55} (1997) 3201; 
J. Hewett and J. Wells, \pr{D55} (1997) 5549;
M. Ciuchini, G. Degrassi, P. Gambino and G.F. Giudice, \np{B534} (1998) 3;
F.M. Borzumati and C. Greub, \pr{D58} (1998) 074004.



\bibitem{msugra} T. Goto, T. Nihei and Y. Okada, \pr{D53} (1996) 5233;
 T. Goto,  Y. Okada, Y. Shimizu and M. Tanaka, \pr{D55} (1997) 4273;
 T. Goto,  Y. Okada and Y. Shimizu, \pr{D58} (1998) 094006.

\bibitem{msugracp} 
S. Baek and P. Ko, KAIST-20/98, hep-ph/9812229;
 T. Goto, Y.-Y. Keum, T. Nihei, Y. Okada and
 Y. Shimizu, KEK-TH-608, hep-ph/9812369.

\bibitem{bkrw} 
A. Pilaftsis and C.E.M. Wagner, hep-ph/9902371, and references therein.

\bibitem{bsg} 
C.-K. Chua, X.-G. He and W.-S. Hou, hep-ph/9808431;
Y.G. Kim, P. Ko and J.S. Lee, hep-ph/9810336;
M. Aoki, G.-C. Cho and N. Oshimo, hep-ph/9811251.

\bibitem{decay} N.G. Deshpande, B. Dutta and S. Oh, \prl{77}, 4499 (1996);
Y. Grossman and M. Worah, \pl{B 395},  241 (1997);
M. Ciuchini \ea, \prl{79}, 978 (1997); 
S.A. Abel, W.N. Cottingham and I.B. Whittingham, \pr{D58} (1998) 073006;
A. Masiero and L. Silvestrini, hep-ph/9709244.

\bibitem{randall} 
L. Randall and S. Su, \np{B540} (1999) 37.

\bibitem{cohen} A.G. Cohen, D.B. Kaplan, F. Lepeintre and A. Nelson, 
\prl{78} (1997) 2300.

\bibitem{barbieri1}
R. Barbieri and A. Strumia, \np{B508} (1997) 3.

\bibitem{grs}
L. Giusti, A. Romanino and A. Strumia, hep-ph/9811386.

\bibitem{oscar}
R. Barbieri, L. Hall, A. Stocchi and N. Weiner, \pl{B425} (1998) 119;
D.A. Demir, A. Masiero and O. Vives, \prl{82} (1999) 2447.


\bibitem{sanda}
A.I. Sanda and Z. Xing, \pr{D56} (1997) 6866;
Z. Xing, \epj{C4} (1998) 283.


\bibitem{lns1} M.~Leurer, Y.~Nir and N.~Seiberg, 
\np{B420} (1994) 468.

\bibitem{ns} Y.~Nir and N.~Seiberg, \pl{B309} (1993) 337.

\bibitem{nr} Y.~Nir and R.~Rattazzi, \pl{B382} (1996) 363.

\bibitem{chmoroi} C.~D.~Carone, L.~J.~Hall and T.~Moroi, 
\pr{D56} (1997) 7183.

\bibitem{bdh} R.~Barbieri, G.~Dvali and L.~J.~Hall, 
\pl{B377} (1996) 76;  R.~Barbieri and L.~J.~Hall, 
Nuovo Cimento {\bf 110} (1997) 1.

\bibitem{bhrr} R.~Barbieri, L.~J.~Hall, S.~Raby and A.~Romanino, 
\np{B493} (1997) 3.

\bibitem{chmu1} C.~D.~Carone, L.~J.~Hall and H.~Murayama, 
\pr{D53} (1996) 6282;  L.~J.~Hall and H.~Murayama, 
\prl{75} (1995) 3985.

\bibitem{chmu2} C.~D.~Carone, L.~J.~Hall and H.~Murayama, 
\pr{D54} (1996) 2328. 

\bibitem{pt} A.~Pomarol and D.~Tommasini, \np{B466} (1996) 3.

\bibitem{dlk} M.~Dine, R.~Leigh and A.~Kagan, \pr{D48} (1993) 4269.

\bibitem{ps} P.~Pouliot and N.~Seiberg, \pl{B318} (1993) 169.

\bibitem{ks} D.~B.~Kaplan and M.~Schmaltz, \pr{D49} (1994) 3741.
 
\bibitem{dg} S.~Dimopoulos and  G.~Giudice, \pl{B357} (1995) 573.

\bibitem{dp} G.~Dvali and A.~Pomarol, \prl{77} (1996) 3728;
G.~Dvali and A.~Pomarol, \np{B522} (1998) 3.

\bibitem{ckn} A.~G.~Cohen, D.~B.~Kaplan and A.~E.~Nelson, 
\pl{B388} (1996) 588.

\bibitem{mr} R.N. Mohapatra and A. Riotto, \pr{D55} (1997) 4262.

\bibitem{nw} A. Nelson and D. Wright, \pr{D56} (1997) 1598.


\bibitem{an} S. Ambrosanio and A. Nelson, \pl{B411} (1997) 283;
S. Ambrosanio and J.D. Wells, hep-ph/9902242.

\bibitem{mi}
L. Hall, V. Kostelecky and S. Raby, \np{B267} (1986) 415.

\bibitem{ggms}
F. Gabbiani, E. Gabrielli, A. Masiero and L. Silvestrini,
\np{B477} (1996) 321.

\bibitem{vm}
J.S. Hagelin, S. Kelley and T. Tanaka, \np{B415} (1994) 293.


\bibitem{akl}
R. Aleksan, B. Kayser and D. London, \prl{73} (1994) 18;
J.P. Silva and L. Wolfenstein, \pr{D55} (1997) 5331;
N.G. Deshpande, B. Dutta and S. Oh, in Ref. \cite{decay}.

\bibitem{al} 
A. Ali and D. London, DESY 99-042, hep-ph/9903535.


\bibitem{ahm}
N. Arkani-Hamed and H. Murayama, \pr{D56} (1997) 6733;
K. Agashe and M. Graesser, \pr{D59} (1998) 015007.

\bibitem{fkp}
J.L. Feng, C. Kolda and N. Polonsky, hep-ph/9810500.


\bibitem{lo}
J.A. Bagger, K.T. Matchev and R.-J. Zhang, \pl{B412} (1997) 77.

\bibitem{nlo}
M. Chiuchini \ea, \np{B523} (1998) 501;
M. Chiuchini \ea, JHEP{\bf 10} (1998) 008;
R. Contino and I. Scimemi, hep-ph/9809437.


\end{thebibliography}
\end{document}